\documentclass[12pt]{article}
\usepackage{graphicx}

\newcommand\pubnumber{NuPhys2016-Mermod}
\newcommand\pubdate{\today}

\def\dpnc{Particle physics department, University of Geneva, Switzerland}

\def\Title#1{\begin{center} {\Large #1 } \end{center}}
\def\Author#1{\begin{center}{ \sc #1} \end{center}}
\def\Address#1{\begin{center}{ \it #1} \end{center}}

\newcommand\pubblock{\rightline{\begin{tabular}{l} \pubnumber\\
         \pubdate  \end{tabular}}}
\newenvironment{Abstract}{\begin{quotation}  }{\end{quotation}}
\newenvironment{Presented}{\begin{quotation} \begin{center} 
             PRESENTED AT\end{center}\bigskip 
      \begin{center}\begin{large}}{\end{large}\end{center} \end{quotation}}

\begin{document}
\begin{titlepage}
\pubblock

\vfill
\Title{Right-handed neutrinos: the hunt is on!}
\vfill

\Author{Philippe Mermod, on behalf of the SHiP Collaboration}
\Address{\dpnc}
\vfill

\begin{Abstract}
The possibility of the existence of right-handed neutrinos remains one of the most important open questions in particle physics, as they can help elucidate the problems of neutrino masses, matter-antimatter asymmetry, and dark matter. Interest in this topic has been increasing in recent years with the proposal of new experimental avenues by which right-handed neutrinos with masses below the electroweak scale could be detected directly using displaced-vertex signatures. At the forefront of such endeavours, the proposed SHiP proton beam-dump experiment is designed for a large acceptance to new weakly-coupled particles and low backgrounds. It is capable of probing right-handed neutrinos with masses below 5~GeV and mixings several orders of magnitude smaller than current constraints, in regions favoured by cosmology. To probe higher masses (up to 30~GeV), a promising novel approach is to identify displaced vertices from right-handed neutrinos produced in $W$ decays at LHC experiments.
\end{Abstract}
\vfill

\begin{Presented}
NuPhys2016, Prospects in Neutrino Physics
Barbican Centre, London, UK,  December 12--14, 2016
\end{Presented}
\vfill

\end{titlepage}
\def\thefootnote{\fnsymbol{footnote}}
\setcounter{footnote}{0}

\section{Introduction}

The phenomenon of neutrino oscillations, showing that neutrinos have masses, provides the first unambiguous microscopic evidence of physics beyond the Standard Model. The electroweak theory was confirmed with the discovery of the Higgs boson in 2012 at the Large Hadron Collider (LHC) and by further measurements of its properties. No signs of new physics have been found so far at the LHC even at its highest collision energies, indicating that, barring neutrino masses, the Standard Model provides a valid description of nature at least up to the TeV scale. It can thus be argued that the greatest challenge faced by experimental particle physics today is to answer the fundamental questions posed by neutrinos: the nature, hierarchy and absolute scale of neutrino masses; the possibility of CP violation in the neutrino sector; and the possible existence of right-handed neutrinos.

\begin{figure}[htb]
\centering
\includegraphics[width=0.49\linewidth]{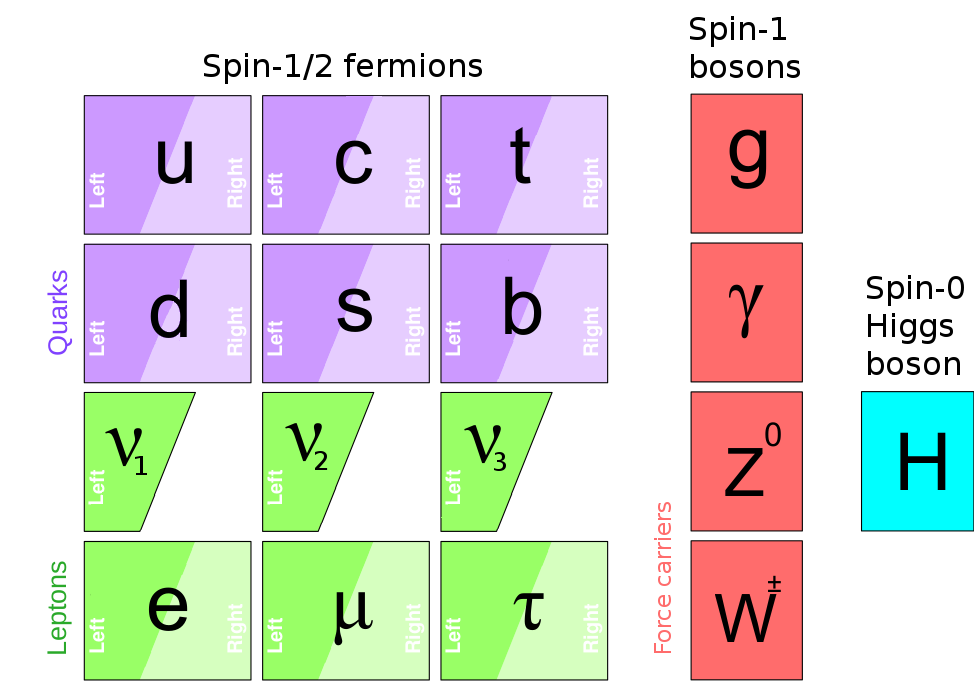}
\includegraphics[width=0.49\linewidth]{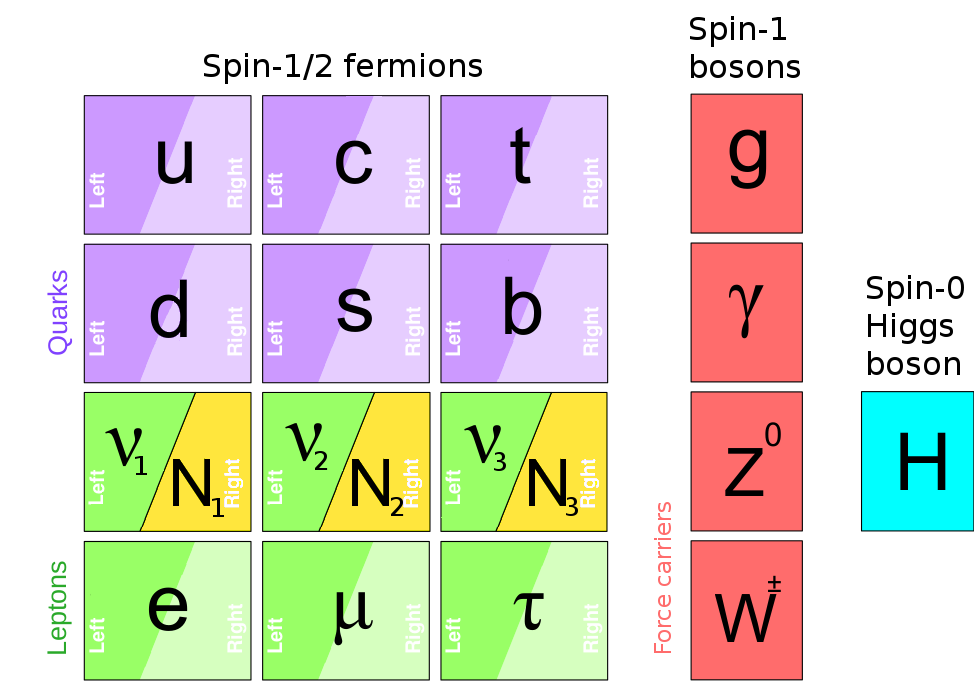}
\caption{Left: summary of the particle states in the Standard Model, indicating left-handed and right-handed fermions separately. All particles in this table have been experimentally observed. Right: three right-handed neutrinos $N_{1,2,3}$ are added and given Majorana masses below the electroweak scale to solve the problems of neutrino masses, dark matter, and matter-antimatter asymmetry in the Universe~\cite{Asaka2005b}. }
\label{fig:nuMSM}
\end{figure}

Remarkably, the hypothesis of three right-handed neutrinos with Majorana masses below the electroweak scale (hereafter termed Heavy Neutral Leptons, or HNLs) in combination with CP violation in the neutrino sector can address at once the three fundamental questions of the origins of neutrino masses, dark matter, and the excess of matter over antimatter in the Universe~\cite{Asaka2005b,Canetti2013a} (see Fig.~\ref{fig:nuMSM}). Being neutral with respect to the electromagnetic, weak, and strong interactions, HNLs are extremely elusive particles which could manifest themselves only through gravitational interactions and by mixing with neutrinos. This mixing is required to obtain small neutrino masses through the seesaw mechanism but needs to be tiny to generate matter-antimatter asymmetry and to evade existing experimental constraints. This means that HNL production in a laboratory would only be possible at the highest beam intensities~\cite{Gorbunov2007}, in addition to the high beam energies required to access high HNL masses. The small mixing also leads to a long lifetime and a typical signature of a displaced decay. 

\section{The SHiP experiment}

A promising strategy to search for HNLs with masses of the order of the GeV is production through hadron decays at high-intensity proton beam-dump facilities. Searches for displaced decays of HNLs with masses up to 0.4~GeV were made using neutrino production through pion and kaon decays, the most sensitive to date with the PS191 experiment at CERN~\cite{Bernardi1988}. Searches were also made using charmed meson decays to access masses up to 2~GeV, the most sensitive to date with the CHARM experiment at CERN~\cite{CHARM1986} and the NuTeV experiment at Fermilab~\cite{NuTeV1999}. For higher HNL masses, up to $75$~GeV, the best constraints come from an analysis of LEP1 data with the DELPHI experiment, where an HNL would be produced in a $Z$ decay and detected as either a prompt or a displaced vertex~\cite{Delphi1997}. 

SHiP is a general-purpose fixed-target facility proposed at the CERN SPS to search for particles with very low couplings to the Standard Model~\cite{Bonivento2013,CERN2014,SHiP2015}. The 400~GeV proton beam extracted from the SPS will be dumped on a high density target with the aim of accumulating $2\times 10^{20}$ protons on target during 5 years of operation. It will produce a large number of neutrinos through hadron decays, following the same principle as that of the CHARM and NuTeV experiments. In particular, neutrinos from decays of hadrons containing $c$ or $b$ quarks can potentially mix with HNLs with masses up to 5~GeV. The charm production at SHiP, with an expected total of $\sim 5\cdot 10^{16}$ neutrinos produced in charm decays, largely surpasses that of any other existing or planned facility, allowing to probe very small coupling strengths and resulting in the HNLs, if produced, to travel very large distances until they decay. 

\begin{figure}[htb]
\centering
\includegraphics[width=0.99\linewidth]{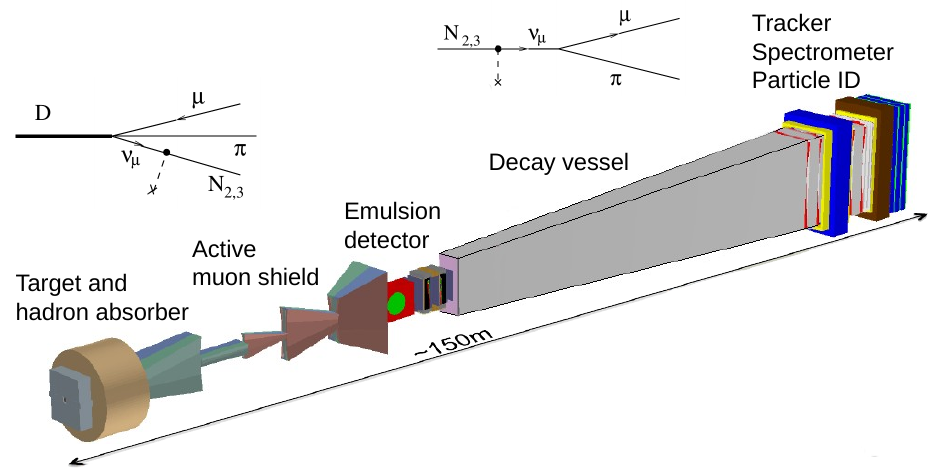}
\caption{Current design of the SHiP experiment. Feynman diagrams of typical processes by which HNLs are produced in the target and decay in the vessel are also shown.}
\label{fig:SHiP}
\end{figure}

In its current design, the SHiP experiment comprises a target followed by a hadron absorber, a muon shield, a 50~m long, 5$-$10~m wide decay volume and a spectrometer similar to the LHCb detector, as shown in Fig.~\ref{fig:SHiP}. The active muon shield is a set of magnets designed to minimise the flux of muons entering the vessel while allowing to have the vessel as close as possible to the target~\cite{SHiP2017a}. The experiment as a whole is optimised to reconstruct and identify decays from new long-lived neutral particles and reject backgrounds which could mimic such decays~\cite{SHiP2015,Alekhin2015}. 

An HNL signal in SHiP is characterised by opposite-sign tracks which cross inside the decay volume, containing at least one particle identified as an electron or a muon, and whose reconstructed parent particle trajectory has its origin at the production target. Backgrounds can possibly arise from neutrino interactions in the vessel, decays of long-lived neutral hadrons, and random crossings of charged particles entering the vessel. Neutrino production in the forward direction is reduced by stopping hadrons in a dense absorber before they decay, and neutrino interactions are minimised by evacuating the air in the vessel. The other types of backgrounds can be vetoed by surrounding the decay volume with tagging detectors. One additional handle to reject random crossings is to measure and match the arrival times of the particles forming the vertex with a high precision ($100$~ps resolution or better) using a dedicated timing detector. Simulations show that the combined use of the active muon shield, veto taggers surrounding the vessel, the timing detector, track momentum and pointing measurements, and muon-pion separation, can reduce the backgrounds to 0.1 events in a sample of $2\times 10^{20}$ protons on target~\cite{SHiP2015}. 

The experimental facility is also ideally suited for studying interactions of tau neutrinos. For this purpose, it will host an emulsion cloud chamber followed by a muon spectrometer upstream of the hidden-particle decay volume.

\section{Heavy neutrino searches at the LHC}

It is generally assumed that new particles accessible at the LHC with masses below 100~GeV would already have been discovered at the LEP, HERA and Tevatron colliders~\cite{PDG2012}. The HNL is an interesting exception. Neutrinos from $W$ and $Z$ decays provide the most efficient way to probe HNLs in the mass range $5-80$~GeV. At previous colliders, they amounted only to a few millions and it was not possible to probe coupling strengths below $10^{-5}$, corresponding to prompt decays for HNL masses above 2.5~GeV. Thus, in previous searches, the vertex displacement could generally not be used as a discriminant against backgrounds, which are important at hadron colliders due to large QCD cross sections. As a consequence, the best current constraints in the HNL mass range $2-75$~GeV come from an analysis with the Delphi experiment at LEP1 using $\sim 10^6$ neutrinos from $Z$ decays~\cite{Delphi1997}. 

\begin{figure}[tb]
\centering
\includegraphics[width=0.6\linewidth]{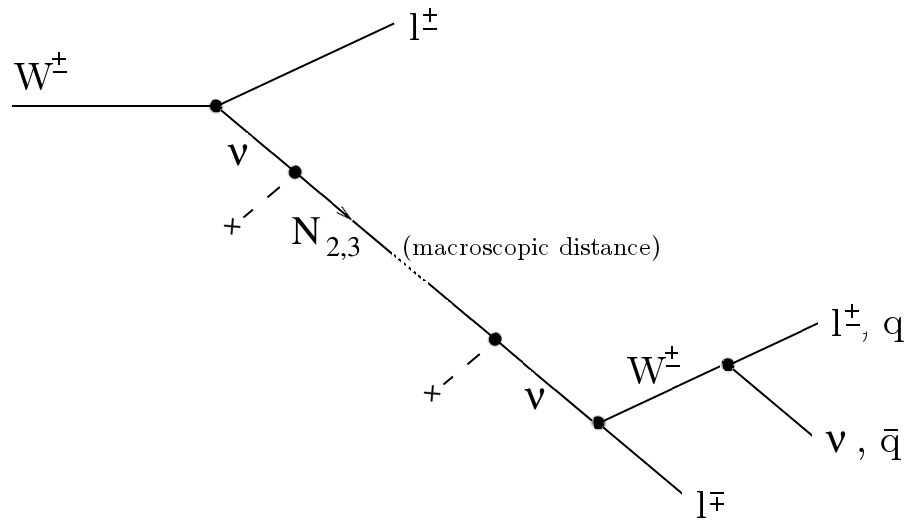}
\caption{Typical process by which a $W$ boson would decay into a long-lived HNL by mixing with a neutrino. The prompt charged lepton is essential for triggering and the displaced vertex allows an efficient background rejection. The leptons from the $W$ decay and the HNL decay can have either opposite or same charge due to the Majorana nature of the HNL, and they can be all three flavours depending on the HNL mixing matrix. The hadronic HNL decay mode (into a charged lepton and two quarks) and the leptonic mode (into two charged leptons and a neutrino) have their respective advantages and should both be considered in the search.}
\label{fig:Wdecay}
\end{figure}

By contrast, a total of $\sim 10^9$ neutrinos from $W$ boson decays were produced at the LHC (ATLAS and CMS) in run-1, and an additional $\sim 10^9$ for every 15~fb$^{-1}$ is being produced in run-2 in 13~TeV collisions since 2016. The process of HNL production through on-shell $W$s and its subsequent decay is illustrated in Fig.~\ref{fig:Wdecay}. It offers two important advantages: the possibility to trigger on the prompt charged lepton from the $W$ decay, and the possibility to efficiently reduce backgrounds by requiring a displaced ($>$few~mm) vertex. Searches using displaced-vertex signatures performed so far at ATLAS~\cite{ATLAS2012b,ATLAS2013c,ATLAS2014a,ATLAS2015b,ATLAS2017a} and CMS~\cite{CMS2014c,CMS2014d,CMS2014b,CMS2016a,CMS2017a} considered the new neutral particles to be decay products of other particles more massive than a $W$, leading to high transverse momentum ($p_T$) displaced decay products. None of these provided any relevant sensitivity to HNLs due to the high-$p_T$ requirements on the particles from the displaced vertex or to the requirement that two displaced vertices should be reconstructed in the same event. However, these searches demonstrate that displaced vertices can be reconstructed with a reasonable efficiency in ATLAS and CMS, and that backgrounds that give rise to such displaced vertices can be kept under control. A well-designed analysis at ATLAS or CMS would be able to probe for the existence of HNLs with masses in the range $3-30$ GeV with a sensitivity which largely surpasses existing LEP constraints and is relevant for BAU generation~\cite{Helo2014,Izaguirre2015}. 

For masses higher than 30~GeV, HNLs produced through on-shell or off-shell $W$ decays would decay promptly. For background reduction one has then to rely on their Majorana nature and require a signature of no opposite-sign same-flavour prompt lepton pairs~\cite{Izaguirre2015}. Searches of this type were made in 8~TeV collisions at ATLAS~\cite{ATLAS2015c} and CMS~\cite{CMS2015b}, setting limits on the HNL mixing to muon neutrinos in the mass range $50-500$~GeV. It should be noted though that the regions of masses and mixings probed using this strategy are not favoured by models of leptogenesis~\cite{Drewes2013}. 

\section{Expected sensitivities}

\begin{figure}[t]
\centering
\includegraphics[width=0.98\linewidth]{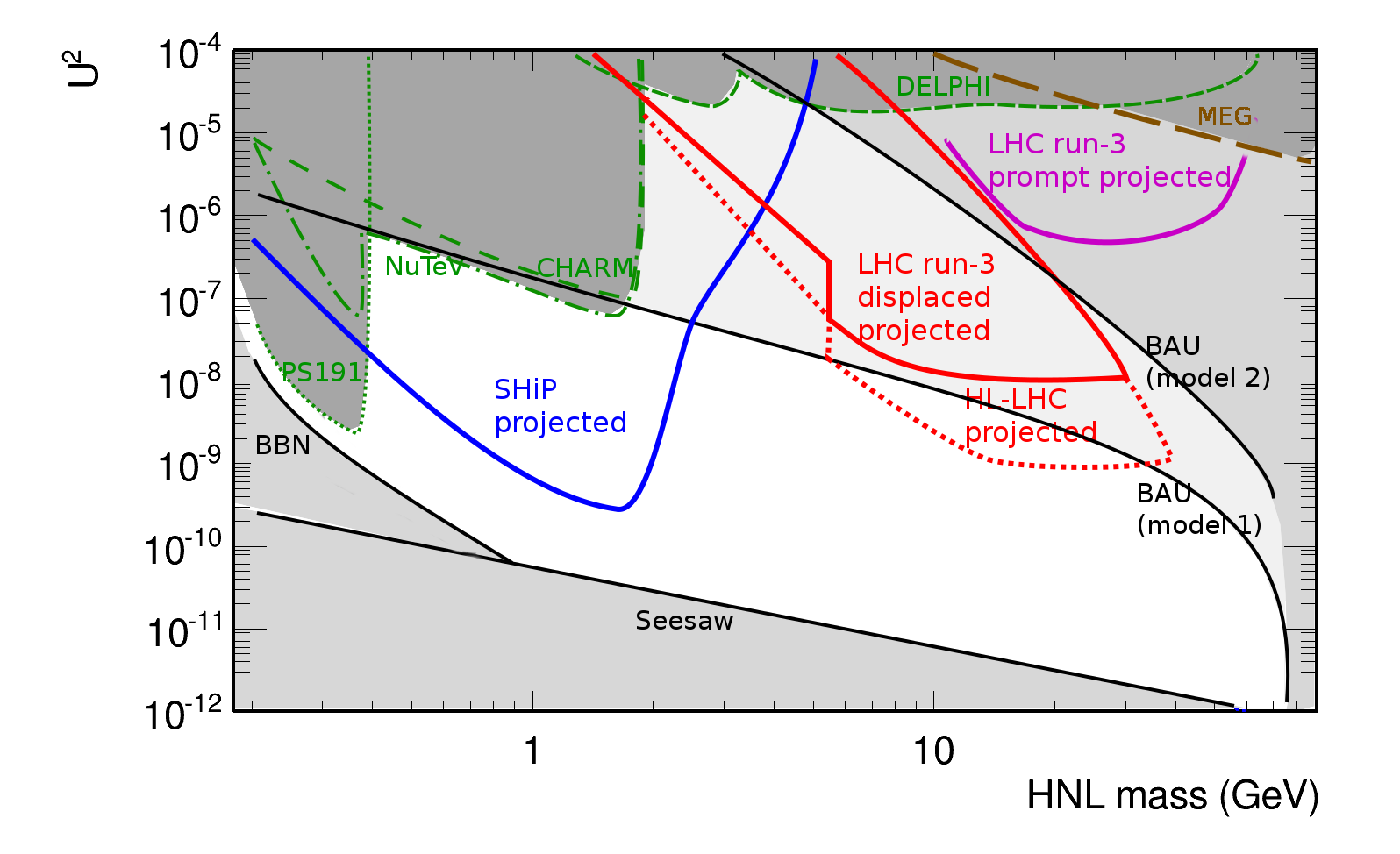}
\caption{Rough expectations for 95\% c.l. exclusion in the HNL coupling strength ($U^2$ with dominant mixing to $\nu_\mu$) vs. mass plane in SHiP and ATLAS/CMS. The specifications of the SHiP technical proposal~\cite{SHiP2015} are assumed. For ATLAS and CMS at run-3 and high-luminosity LHC, 300 fb$^{-1}$ and 3000 fb$^{-1}$ of 14 TeV proton-proton collisions are assumed, respectively. $U^2$ is constrained from below by the observed neutrino masses and by primordial nucleosynthesis (BBN)~\cite{Canetti2013b}. Upper constraints on $U^2$ are shown for two different models accounting for baryon asymmetry in the Universe (BAU): model 1 uses one HNL for dark matter~\cite{Canetti2013b} while model 2 requires all three HNLs to contribute to BAU~\cite{Canetti2014}. Direct~\cite{Bernardi1988,CHARM1986,NuTeV1999,Delphi1997,CMS2015b} and indirect~\cite{MEG2013,Antusch2015} experimental limits are indicated as dashed green and brown lines, respectively.}
\label{fig:sensitivity}
\end{figure}

Assuming the layout of the SHiP experiment which is described in the technical proposal~\cite{SHiP2015}, with $2\times 10^{20}$ protons on target and acceptance and backgrounds with a basic event selection estimated from simulations, one obtains the sensitivity curve shown in blue in Fig.~\ref{fig:sensitivity}. This is only a preliminary estimate as the experiment is still being optimised for a trade-off between cost and performance, but it indicates clearly that SHiP will be able to dig deeply into the most favoured regions of the parameter space. 

The curves shown in red and purple in Fig.~\ref{fig:sensitivity} show rough expected sensitivities of dedicated ATLAS and CMS analyses using the full LHC run-3 dataset with a signature of a displaced vertex or prompt no opposite-charge same-sign leptons, respectively. For the displaced-vertex signature, both leptonic and hadronic HNL decays are considered, with a reconstructed vertex mass cut set to $>4.5$~GeV for the hadronic channel to reduce hadronic backgrounds. Other assumptions made are: zero background, 60\% trigger efficiency, 20\% vertex selection efficiency, and 50\% efficiency for the final selection cuts. For the prompt analysis, the same assumptions as in Ref.~\cite{Izaguirre2015} are made. While the efficiency and background assumptions still need to be confirmed with detailed studies, it is clear that dedicated HNL searches at the LHC will largely surpass the existing LEP constraints in the mass range $3-50$ GeV. 

\section{Summary}

The proposed SHiP experiment plans to use the CERN SPS high-intensity proton beam to probe the smallest possible HNL mixings to neutrinos -- several orders of magnitude smaller than current limits in the $0.4-3$~GeV HNL mass range. SHiP is widely recognised as an important complement to existing CERN programmes after LHC run-2. In a complementary manner, displaced-vertex signatures at ATLAS and CMS allow to access higher HNL masses thanks to a high rate of $W$ boson production. These two approaches offer unique opportunities to probe the existence of right-handed neutrinos below the electroweak scale by direct production and detection at the CERN laboratory within a 10-year time scale. The reward is potentially very high, as such particles can shed light on the mechanisms behind the observed neutrino masses and the generation of baryon asymmetry in the Universe. 



\bibliographystyle{mystylem}
\bibliography{HNLrefs}

\providecommand{\href}[2]{#2}\begingroup\raggedright\begin{thebibliography}{10}

\bibitem{Asaka2005b}
T.~Asaka and M.~Shaposhnikov, {\em {The nuMSM, dark matter and baryon asymmetry
  of the universe}\/},
  \href{http://dx.doi.org/10.1016/j.physletb.2005.06.020}{Phys. Lett. B {\bf
  620} (2005)  17}, \href{http://arxiv.org/abs/hep-ph/0505013}{{\tt
  arXiv:hep-ph/0505013 [hep-ph]}}.

\bibitem{Canetti2013a}
L.~Canetti, M.~Drewes, and M.~Shaposhnikov, {\em {Sterile Neutrinos as the
  Origin of Dark and Baryonic Matter}\/},
  \href{http://dx.doi.org/10.1103/PhysRevLett.110.061801}{Phys. Rev. Lett. {\bf
  110} (2013)  061801}, \href{http://arxiv.org/abs/1204.3902}{{\tt
  arXiv:1204.3902 [hep-ph]}}.

\bibitem{Gorbunov2007}
D.~Gorbunov and M.~Shaposhnikov, {\em {How to find neutral leptons of the
  $\nu$MSM?}\/},  \href{http://dx.doi.org/10.1007/JHEP11(2013)101,
  10.1088/1126-6708/2007/10/015}{JHEP {\bf 0710} (2007)  015},
  \href{http://arxiv.org/abs/0705.1729}{{\tt arXiv:0705.1729 [hep-ph]}}.

\bibitem{Bernardi1988}
G.~Bernardi, G.~Carugno, J.~Chauveau, F.~Dicarlo, M.~Dris, J.~Dumarchez,
  M.~Ferro-Luzzi, J.-M. Levy, D.~Lukas, and J.-M. Perreau, {\em {Further limits
  on heavy neutrino couplings}\/},
  \href{http://dx.doi.org/10.1016/0370-2693(88)90563-1}{Phys. Lett. B {\bf 203}
  (1988)  332}.

\bibitem{CHARM1986}
{CHARM Collaboration}, {\em {A Search for Decays of Heavy Neutrinos in the Mass
  Range 0.5-{GeV} to 2.8-{GeV}}\/},
  \href{http://dx.doi.org/10.1016/0370-2693(86)91601-1}{Phys. Lett. B {\bf 166}
  (1986)  473}.

\bibitem{NuTeV1999}
{NuTeV Collaboration}, {\em {Search for neutral heavy leptons in a high-energy
  neutrino beam}\/},
  \href{http://dx.doi.org/10.1103/PhysRevLett.83.4943}{Phys. Rev. Lett. {\bf
  83} (1999)  4943}, \href{http://arxiv.org/abs/hep-ex/9908011}{{\tt
  arXiv:hep-ex/9908011 [hep-ex]}}.

\bibitem{Delphi1997}
{DELPHI Collaboration}, {\em {Search for neutral heavy leptons produced in Z
  decays}\/},  \href{http://dx.doi.org/10.1007/s002880050370}{Z. Phys. C {\bf
  74} (1997)  57}.

\bibitem{Bonivento2013}
W.~Bonivento, A.~Boyarsky, H.~Dijkstra, U.~Egede, M.~Ferro-Luzzi, B.~Goddard,
  A.~Golutvin, D.~Gorbunov, R.~Jacobsson, J.~Panman, M.~Patel, O.~Ruchayskiy,
  T.~Ruf, N.~Serra, M.~Shaposhnikov, and D.~Treille, {\em {Proposal to Search
  for Heavy Neutral Leptons at the SPS}\/},  CERN-SPSC-2013-024, SPSC-EOI-010
  (2013)  , \href{http://arxiv.org/abs/1310.1762}{{\tt arXiv:1310.1762
  [hep-ex]}}.

\bibitem{CERN2014}
{CERN task force report}, {\em {A new Experiment to Search for Hidden Particles
  (SHIP) at the SPS North Area -- Preliminary Project and Cost Estimate}\/},
  EN-DH-2014-007 (2014)  .

\bibitem{SHiP2015}
{SHiP Collaboration}, {\em {A facility to Search for Hidden Particles (SHiP) at
  the CERN SPS}\/},  CERN-SPSC-2015-016 (2015)  ,
  \href{http://arxiv.org/abs/1504.04956}{{\tt arXiv:1504.04956
  [physics.ins-det]}}.

\bibitem{SHiP2017a}
{SHiP Collaboration}, {\em {The active muon shield in the SHiP experiment}\/},
  accepted for publication in JINST (2017)  ,
  \href{http://arxiv.org/abs/1703.03612}{{\tt arXiv:1703.03612
  [physics.ins-det]}}.

\bibitem{Alekhin2015}
S.~Alekhin et al., {\em {A facility to Search for Hidden Particles at the CERN
  SPS: the SHiP physics case}\/},  CERN-SPSC-2015-017 (2015)  ,
  \href{http://arxiv.org/abs/1504.04855}{{\tt arXiv:1504.04855 [hep-ph]}}.

\bibitem{PDG2012}
{Particle Data Group} Collaboration, J.~Beringer et al., {\em {Review of
  Particle Physics (RPP)}\/},
  \href{http://dx.doi.org/10.1103/PhysRevD.86.010001}{Phys. Rev. D {\bf 86}
  (2012)  010001}.

\bibitem{ATLAS2012b}
{ATLAS Collaboration}, {\em {Search for a light Higgs boson decaying to
  long-lived weakly-interacting particles in proton-proton collisions at
  $\sqrt{s}=7$ TeV with the ATLAS detector}\/},
  \href{http://dx.doi.org/10.1103/PhysRevLett.108.251801}{Phys. Rev. Lett. {\bf
  108} (2012)  251801}, \href{http://arxiv.org/abs/1203.1303}{{\tt
  arXiv:1203.1303 [hep-ex]}}.

\bibitem{ATLAS2013c}
{ATLAS Collaboration}, {\em {Search for displaced muonic lepton jets from light
  Higgs boson decay in proton-proton collisions at $\sqrt{s}=7$ TeV with the
  ATLAS detector}\/},
  \href{http://dx.doi.org/10.1016/j.physletb.2013.02.058}{Phys. Lett. B {\bf
  721} (2013)  32}, \href{http://arxiv.org/abs/1210.0435}{{\tt arXiv:1210.0435
  [hep-ex]}}.

\bibitem{ATLAS2014a}
{ATLAS Collaboration}, {\em {Search for long-lived neutral particles decaying
  into lepton jets in proton-proton collisions at $\sqrt{s}=8 $ TeV with the
  ATLAS detector}\/},  \href{http://dx.doi.org/10.1007/JHEP11(2014)088}{JHEP
  {\bf 1411} (2014)  088}, \href{http://arxiv.org/abs/1409.0746}{{\tt
  arXiv:1409.0746 [hep-ex]}}.

\bibitem{ATLAS2015b}
{ATLAS Collaboration}, {\em {Search for massive, long-lived particles using
  multitrack displaced vertices or displaced lepton pairs in pp collisions at
  $\sqrt{s}$ = 8 TeV with the ATLAS detector}\/},  Phys. Rev. D {\bf 92} (2015)
   072004, \href{http://arxiv.org/abs/1504.05162}{{\tt arXiv:1504.05162
  [hep-ex]}}.

\bibitem{ATLAS2017a}
{ATLAS Collaboration}, {\em {Search for long-lived, massive particles in events
  with displaced vertices and missing transverse momentum in 13 TeV $pp$
  collisions with the ATLAS detector}\/},  ATLAS-CONF-2017-026 (2017)  .

\bibitem{CMS2014c}
{CMS Collaboration}, {\em {Search for long-lived particles that decay into
  final states containing two electrons or two muons in proton-proton
  collisions at sqrt(s) = 8 TeV}\/},  Phys. Rev. D {\bf 91} (2015)  052012,
  \href{http://arxiv.org/abs/1411.6977}{{\tt arXiv:1411.6977 [hep-ex]}}.

\bibitem{CMS2014d}
{CMS Collaboration}, {\em {Search for long-lived neutral particles decaying to
  quark-antiquark pairs in proton-proton collisions at sqrt(s) = 8 TeV}\/},
  Phys. Rev. D {\bf 91} (2015)  012007,
  \href{http://arxiv.org/abs/1411.6530}{{\tt arXiv:1411.6530 [hep-ex]}}.

\bibitem{CMS2014b}
{CMS Collaboration}, {\em {Search for "Displaced Supersymmetry" in events with
  an electron and a muon with large impact parameters}\/},  Phys. Rev. Lett.
  {\bf 114} (2015)  061801, \href{http://arxiv.org/abs/1409.4789}{{\tt
  arXiv:1409.4789 [hep-ex]}}.

\bibitem{CMS2016a}
{CMS Collaboration}, {\em {Search for displaced leptons in the e-mu
  channel}\/},  CMS-PAS-EXO-16-022 (2016)  .

\bibitem{CMS2017a}
{CMS Collaboration}, {\em {Inclusive search for new particles decaying to
  displaced jets at $\sqrt{s} = 13~\mathrm{TeV}$}\/},  {CMS-PAS-EXO-16-003}
  (2017)  .

\bibitem{Helo2014}
J.~Helo, M.~Hirsch, and S.~Kovalenko, {\em {Heavy neutrino searches at the LHC
  with displaced vertices}\/},
  \href{http://dx.doi.org/10.1103/PhysRevD.89.073005}{Phys. Rev. D {\bf 89}
  (2014)  073005}, \href{http://arxiv.org/abs/1312.2900}{{\tt arXiv:1312.2900
  [hep-ph]}}.

\bibitem{Izaguirre2015}
E.~Izaguirre and B.~Shuve, {\em {Multilepton and Lepton Jet Probes of
  Sub-Weak-Scale Right-Handed Neutrinos}\/},
  \href{http://dx.doi.org/10.1103/PhysRevD.91.093010}{Phys. Rev. D {\bf 91}
  (2015) no.~9, 093010}, \href{http://arxiv.org/abs/1504.02470}{{\tt
  arXiv:1504.02470 [hep-ph]}}.

\bibitem{ATLAS2015c}
{ATLAS Collaboration}, {\em {Search for heavy Majorana neutrinos with the ATLAS
  detector in pp collisions at $ \sqrt{s}=8 $ TeV}\/},
  \href{http://dx.doi.org/10.1007/JHEP07(2015)162}{JHEP {\bf 07} (2015)  162},
  \href{http://arxiv.org/abs/1506.06020}{{\tt arXiv:1506.06020 [hep-ex]}}.

\bibitem{CMS2015b}
{CMS Collaboration}, {\em {Search for heavy Majorana neutrinos in $\mu^\pm
  \mu^\pm+$ jets events in proton-proton collisions at $\sqrt{s}$ = 8 TeV}\/},
  \href{http://dx.doi.org/10.1016/j.physletb.2015.06.070}{Phys. Lett. B {\bf
  748} (2015)  144}, \href{http://arxiv.org/abs/1501.05566}{{\tt
  arXiv:1501.05566 [hep-ex]}}.

\bibitem{Drewes2013}
M.~Drewes, {\em {The Phenomenology of Right Handed Neutrinos}\/},
  \href{http://dx.doi.org/10.1142/S0218301313300191}{Int. J. Mod. Phys. E {\bf
  22} (2013)  1330019}, \href{http://arxiv.org/abs/1303.6912}{{\tt
  arXiv:1303.6912 [hep-ph]}}.

\bibitem{Canetti2013b}
L.~Canetti, M.~Drewes, T.~Frossard, and M.~Shaposhnikov, {\em {Dark Matter,
  Baryogenesis and Neutrino Oscillations from Right Handed Neutrinos}\/},
  \href{http://dx.doi.org/10.1103/PhysRevD.87.093006}{Phys. Rev. D {\bf 87}
  (2013)  093006}, \href{http://arxiv.org/abs/1208.4607}{{\tt arXiv:1208.4607
  [hep-ph]}}.

\bibitem{Canetti2014}
L.~Canetti, M.~Drewes, and B.~Garbrecht, {\em {Probing leptogenesis with
  GeV-scale sterile neutrinos at LHCb and Belle II}\/},
  \href{http://dx.doi.org/10.1103/PhysRevD.90.125005}{Phys. Rev. D {\bf 90}
  (2014)  125005}, \href{http://arxiv.org/abs/1404.7114}{{\tt arXiv:1404.7114
  [hep-ph]}}.

\bibitem{MEG2013}
{MEG Collaboration}, {\em {New constraint on the existence of the $\mu^+ \to
  e^+\gamma$ decay}\/},
  \href{http://dx.doi.org/10.1103/PhysRevLett.110.201801}{Phys. Rev. Lett. {\bf
  110} (2013)  201801}, \href{http://arxiv.org/abs/1303.0754}{{\tt
  arXiv:1303.0754 [hep-ex]}}.

\bibitem{Antusch2015}
S.~Antusch and O.~Fischer, {\em {Testing sterile neutrino extensions of the
  Standard Model at future lepton colliders}\/},
  \href{http://dx.doi.org/10.1007/JHEP05(2015)053}{JHEP {\bf 05} (2015)  053},
  \href{http://arxiv.org/abs/1502.05915}{{\tt arXiv:1502.05915 [hep-ph]}}.

\end{thebibliography}\endgroup

%
%
%
%
%
%
%
%

\end{document}